\documentclass[10pt,a4paper]{article}
\usepackage{bm}  %% bold math
\usepackage{amsmath}  %% align, \eqref, \dddot

\newcommand\nc{\newcommand*}  \nc\longnc{\newcommand}

\nc\nn{\tag*{}}  %% line with no number in align
\nc\Lit[2]{{#1}_{\{#2\}}}  %% #2: which iteration of #1
\nc\LLit[4]{\hbox{${#1}_     %% #2: which iteration of #1,
  {\{#2\}}^{\,\,#3}$}_{#4}}  %% #3: superscript, #4: subscript
\nc\LK[1]{{\hat{#1}}}  %% quantity per mass
\nc\LH[1]{{\tilde{#1}}}  %% quantity times e/m
\nc\fracc[2]{#1/#2}  \nc\fraccc[2]{#1/\left(#2\right)}
\nc\three{\bm}  %% three-vectors and -tensors
\nc\LLnull{\three{0}}  %% the zero three-vector
\nc\cc{\mathrm{c.\,c.}}  %% complex conjugate
\nc\Lcdot{\,\cdot\,}
\nc\dd{\mathrm{d}}  %% differential
\nc\e{\mathrm{e}}  %% 2.718281828459...
\nc\ii{\mathrm{i}}  %% the imaginary unit
\nc\qtraf[1]{{#1}^{\LP}}

\nc\LLA{\LLK}
\nc\LLB{\three{F}}
\nc\LLHB{\LH{\LLB}}
\nc\LF{F}
\nc\LLI{\three{I}}
\nc\LLK{\three{L}}
\nc\LL{L}
\nc\LP{P}
\nc\LLP{\three{P}}
\nc\LLT{\three{T}}
\nc\LLW{\three{W}}

\nc\LLLa{\Lx_0}
\nc\LLLLa{c}
\nc\Lb{B}  \nc\LHb{\LH{\Lb}}  \nc\LLb{\three{\Lb}}  \nc\LLHb{\LH{\LLb}}
\nc\LLLLb{z}
\nc\Lc{c}
\nc\LLc{\LLv_0}
\nc\LLLc{z_0}
\nc\Le{E}  \nc\LHe{\LH{\Le}}  \nc\LLe{\three{\Le}}  \nc\LLHe{\LH{\LLe}}
\nc\Lf{f}  \nc\LKf{\LK{\Lf}}  \nc\LLf{\three{\Lf}}  \nc\LLKf{\LK{\LLf}}
\nc\LLLf{p}
\nc\LLLg{q}
\nc\Lg{g}
\nc\LLh{\three{h}}
\nc\LLLh{\Lv_0}
\nc\Lm{m}
\nc\Lfs{f_{\mathrm{self}}}
\nc\Lp{p}
\nc\LLp{p}
\nc\LLq{q}
\nc\Lq{e}
\nc\Lrr{r}
\nc\Lr{x}  \nc\LLr{\three{\Lr}}  \nc\Lrnull{x_0}
\nc\Lrdotnull{u_0}
\nc\Ls{b}  \nc\LKs{\LK{\Ls}}  \nc\LLs{{\three{\Ls}}}  \nc\LLKs{\LK{\LLs}}
\nc\LLLs{s}
\nc\Lt{t}
\nc\Lu{u}
\nc\LLu{\three{u}}
\nc\Lv{v}
\nc\LLv{\three{v}}
\nc\LLLv{\three{w}}
\nc\LLw{\three{w}}
\nc\Lx{x}
\nc\LLx{x}
\nc\Lz{r}

\nc\Lalp{\gamma}
\nc\Lbet{\alpha}  %% finomszerk. allando...
\nc\Lphi{\varrho}
\nc\Llam{\lambda}
\nc\LLlam{\kappa}
\nc\LLLlam{\beta}
\nc\Lmu{\mu}
\nc\Lnu{\nu}
\nc\Lxi{\xi}
\nc\Leta{\eta}  \nc\LKeta{\LK{\Leta}}
\nc\Lome{\omega}
\nc\Lpi{\zeta}
\nc\Lro{\rho}

\begin{document}

 \begin{center}
  \LARGE\bf
Second order equation of motion for electromagnetic radiation back-reaction
 \end{center}

\footnotetext{\hspace*{-.45cm}\footnotesize
$^*$Corresponding author. E-mail: fulop@energia.bme.hu}

\begin{center}
\rm T. Matolcsi$^{\rm a)}$, \ \ T. F\"ul\"op$^{\rm b)*}$, \ and
\ M. Weiner$^{\rm c)}$
\end{center}

 \begin{center}
 \begin{footnotesize} \sl
${}^{\rm a)}$ Department of Applied Analysis and Computational Mathematics,
E\"otv\"os Lor\'and University,
 \\
P\'azm\'any P.~stny.~1/C, H-1117 Budapest, Hungary
 \\ 
${}^{\rm b)}$ Department of Energy Engineering,
Budapest University of Technology and Economics,
 \\
Bertalan L.~u.~4-6, H-1111 Budapest, Hungary
 \\
${}^{\rm c)}$ Department of Analysis,
Budapest University of Technology and Economics,
 \\
Egry J.~u.~1, H-1111 Budapest, Hungary
 \\
 \end{footnotesize}
 \end{center}

\vskip2ex

\begin{abstract}
We take the viewpoint that the physically acceptable solutions of the
Lorentz--Dirac equation for radiation back-reaction are actually
determined by a second order equation of motion, the self-force being
given as a function of spacetime location and velocity. We propose three
different methods to obtain this self-force function. For two example
systems, we determine the second order equation of motion exactly in the
nonrelativistic regime via each of these three methods, the three
methods leading to the same result. We reveal that, for both systems
considered, back-reaction induces a damping proportional to velocity
and, in addition, it decreases the effect of the external force.
%\keywords{electromagnetic radiation; back-reaction; Lorentz--Dirac equation.}
\end{abstract}

%\ccode{PACS Nos.: 03.50.De, 41.60.-m, 45.20.D-, 11.10.Gh}

\section{Introduction}

The Maxwell equations of classical electrodynamics describe how {%\it
given} charged particles determine the electromagnetic field, while the
Newton equation with the Lorentz force tells how a {%\it
given}
electromagnetic field determines the motion of charged particles. As a
first step towards describing {%\it
interaction} between matter and
field, the radiation back-reaction (self-force) of a point charge $\Lq$
having the special relativistic world line function $\Lrr$ is deduced to
be\cite{Jackson,%
%} Ch.~16.3, \cite{
deGroot,%
%} Ch.~III.3, \cite{
Taylor,Matolcsi,Rowe,Gsponer}
 \begin{equation}  \label{self}
\Lfs^j = \Leta \left( {\Lg^j}_k - \fracc{\dot{\Lrr}^j \dot{\Lrr}_k}{\Lc^2}
\right) \dddot \Lrr^k ,
 \end{equation}
where indices run from 0 to 3, $\Lg$ is the spacetime metric, $\Lc$ is
the speed of light, $\Leta = (2/3) \, \Lq^2/\Lc^3$, and overdot denotes
differentiation with respect to proper time%
%\footnote{
\footnotetext{\hspace*{-.45cm}$^1$\footnotesize
Note that, upon
$\dot{\Lrr}^j \dot{\Lrr}_j = 1$, we also have $- \dot{\Lrr}^j
\dot{\Lrr}_k \dddot\Lrr^k = \dot{\Lrr}^j \left[ \ddot{\Lrr}_k
\ddot\Lrr^k - ({1}/{2}) \big( \dot{\Lrr}^k \dot{\Lrr}_k \big)
\ddot{\vphantom{|}} \; \right] = \dot{\Lrr}^j \ddot{\Lrr}_k
\ddot\Lrr^k$. Frequently, the self-force is written using this latter
form.}.$^1$\addtocounter{footnote}{1}
 This force is added to the external force $\Lf$, which may depend on
both the spacetime location and the four-velocity of the particle, to
obtain
 \begin{equation}  \label{mot}
\Lm \ddot \Lrr^j = \Lf^{j} (\Lrr, \dot{\Lrr}) + \Leta \left( {\Lg^j}_k -
\fracc{\dot{\Lrr}^j \dot{\Lrr}_k}{\Lc^2} \right) \dddot \Lrr^k ,
 \end{equation}
called the Lorentz--Dirac equation, for the motion of the point particle
with mass $\Lm$.

The problems with this equation are well-known. First, it is of third
order so the initial values of spacetime position, velocity and
acceleration are necessary to obtain the motion, and there is no
apparent reasoning how to prescribe acceleration. Second, the equation
admits 'runaway' solutions---motions accelerating exponentially in
time---, and, third, it exhibits acausal behavior.

There are a number of attempts to treat these problems (see, e.g.,
%Refs.~\ref
\cite{Dirac,Haag,deSouza,Spohn00,OConnell,Spohn04,%
Yaghjian,Rohrlich07,Mares,Parga,Kar}, research along the lines of
%Ref.~\ref
\cite{ForHerKov}, etc.), trying to find the physically acceptable
solutions of the Lorentz--Dirac equation and giving further and further
insight into the situation.

It is important to note here that, irrespective of whether the
Lorentz--Dirac equation is considered an exact one for exactly
pointlike charges or an approximate one for distributed ones,
% the question is not necessarily
it is legitimate, useful and insightful to investigate whether an
equation of the form \eqref{mot} finds physically meaningful use.
%can serve as a physically acceptable one.

One of the proposed and most frequently applied approach is as
follows\cite{Landau}. As a zeroth approximation, the equation without
radiation is considered:
 \begin{equation}  \label{qek}
\Lm \ddot \Lr^j = \Lf^{j} (\Lr, \dot{\Lr}) .
 \end{equation}
The third derivative is computed from this equation,
 \begin{equation}  \label{dot3}
\dddot{\Lr}^j = \frac{1}{\Lm} \left( \frac{\partial \Lf^{j}}{\partial
\Lr^k} \dot{\Lr}^k + \frac{\partial \Lf^{j}}{\partial \dot\Lr^k}
\ddot{\Lr}^k \right) .
 \end{equation}
Then the second derivative here is replaced by the rhs of \eqref{qek}
(divided by $\Lm$), an expression of lower order derivatives, leading to
the following approximation of the self-force \eqref{self}:
 \begin{equation}  \label{qem}
\LLit{\Ls}{1}{j}{} ( \Lr, \dot{\Lr} ) := \Leta \left( {\Lg^j}_k -
\frac{\dot{\Lr}^j \dot{\Lr}_k}{\Lc^2} \right) \frac{1}{\Lm} \left(
\frac{\partial \Lf^{k}}{\partial \Lr^l} \dot{\Lr}^l + \frac{1}{\Lm}
\frac{\partial \Lf^{k}}{\partial \dot\Lr^l} \Lf^{l} ( \Lr, \dot{\Lr} )
\right) .
 \end{equation}
This is added to the external force to derive an approximate second
order equation of motion:
 \begin{equation}  \label{mot1}
\Lm \ddot{\Lr}^j = \Lf^{j} ( \Lr, \dot{\Lr} ) + \LLit{\Ls}{1}{j}{} (
\Lr, \dot{\Lr} ) .
 \end{equation}
In known examples, $\LLit{\Ls}{1}{j}{} ( \Lr, \dot{\Lr} )$ is found to
be a damping dissipative term.

Another idea \cite{Spohn00} is that the initial values of spacetime
position, velocity and acceleration cannot be given independently, and
one has to find a `critical manifold' formed by those initial values
which do not result in runaway solutions. In
% Ref.~\ref
\cite{Spohn00}, it is
stated that the critical manifold admits a second order differential
equation for the physically acceptable motions but the actual form of
such an equation is not given. Instead, by a singular perturbation, only
a first approximation is provided, which results in \eqref{mot1}.

Here, our object of interest is the exact form of the second order
differential equation for the critical manifold.
Before proceeding, we mention that other results \cite{Pol} also
suggest, implicitly, the existence of a second order equation of motion
in the background.

In what follows, we explicitly assume the existence of a second order
equation of motion,%
%\footnote{
\footnotetext{\hspace*{-.45cm}$^2$\footnotesize
Apparently, this hypothesis is considerably
stronger than assuming dependence on the extended past, which phenomenon
could also be plausible by physical expectations about how interaction
takes place between matter and field. Here, driven by the above
motivations, we investigate whether one can succeed with this stronger
assumption.}$^2$\addtocounter{footnote}{1}
 \begin{equation}  \label{mott}
\Lm \ddot{\Lr}^j = \Lf^{j} ( \Lr, \dot{\Lr} ) + \Ls^j ( \Lr, \dot\Lr ) ,
 \end{equation}
with the self-force as a function of spacetime position and velocity,
 \begin{equation}  \label{selft}
\Lfs^j = \Ls^j ( \Lr, \dot\Lr ),
 \end{equation}
and derive the condition on it that ensures that all its solutions are
solutions of \eqref{mot} as well.  This condition turns out to take the
form of a partial differential equation for \eqref{selft}.

In parallel, based on physically plausible ideas, we also propose two
iterative methods for obtaining the self-force function. Actually, the
first step of one of these iterative methods corresponds to
\eqref{mot1}.

For demonstration, we investigate the partial differential equation and
the iterative methods in quantitative detail, in the nonrelativistic
regime,%
%\footnote{
\footnotetext{\hspace*{-.45cm}$^3$\footnotesize
Namely, we take only $j = 1, 2, 3$ in \eqref{mot} and
omit terms of the order of ${1}/{\Lc^2}$ or higher.}$^3$\addtocounter{footnote}{1}
% \cite{nonrel}
treating two special cases. For both systems, we calculate the   
self-force function exactly via each of the three methods, and find that
the three approaches provide the same result.

\section{The self-force function}

For convenience, we introduce the shorthands
 \begin{equation}  \label{qez}
%\LKf := \frac{1}{\Lm} \Lf ,  \qquad  \LKeta := \frac{1}{\Lm} \Leta ,
% \qquad  \LKs := \frac{1}{\Lm} \Ls .
\LKf := \Lf / \Lm ,
 \qquad
\LKeta := \Leta / \Lm,
 \qquad
\LKs := \Ls / \Lm .
 \end{equation}

\subsection{Differential equation for the self-force function}  \label{force}

As said, we assume that the equation of motion of a radiating particle
is of the form
 \begin{equation}  \label{qav}
\ddot{\Lr}^j = \LKf^{j} (\Lr, \dot{\Lr}) + \LKs^j (\Lr, \dot{\Lr}) ,
 \end{equation}
where the second term on the rhs is the self-force as a function of
spacetime position and four-velocity. In order to obtain its actual
expression, we use a fixed-point-like property, as follows. Computing
the third derivative from the expected equation of motion \eqref{qav},
substituting $\ddot \Lr$ in the obtained expression by the rhs of
\eqref{qav}, and applying $\LKeta \left( {\Lg^j}_k - {\dot{\Lr}^j
\dot{\Lr}_k}/{\Lc^2} \right)$ to the result, we have to recover the
self-force:
 \begin{align}  \label{qen}
\LKs^j
% &
=
\LKeta \left( {\Lg^j}_k - \frac{\dot{\Lr}^j \dot{\Lr}_k}{\Lc^2} \right)
% \nn\\   \label{qgc}
% &
% \quad
% \times
\left[ \frac{\partial \bigl( \LKf^k + \LKs^k \bigr) }{\partial
\Lr^l} \dot{\Lr}^l + \frac{\partial \bigl( \LKf^k + \LKs^k \bigr)
}{\partial \dot\Lr^l} \bigl( \LKf^l + \LKs^l \bigr) \right] .
 \end{align}
Naturally, one has to keep in mind that
%, here,
 $\Lr$ and $\dot \Lr$ are
to be understood as independent variables. We can display this fact in a
more self-explaining way, writing \eqref{qen} as
 \begin{align}  \label{qfn}
\LKs^j (\Lr, \Lu)
 &
=
\LKeta \left( {\Lg^j}_k - \frac{\Lu^j \Lu_k}{\Lc^2}
\right)
% \nn\\   \label{qgb}
% &
%\hskip 1em
%\times
\left\{ \frac{\partial \bigl[ \LKf^k + \LKs^k \bigr] (\Lr, \Lu)
}{\partial \Lr^l} \Lu^l + \frac{\partial \bigl[ \LKf^k + \LKs^k \bigr]
(\Lr, \Lu)}{\partial \Lu^l} \bigl[ \LKf^l + \LKs^l \bigr] (\Lr, \Lu)
\right\} .
 \end{align}

This is a first order partial differential equation (system) for the
two-variable function(s) $\LKs^j$. Its solution is expected to contain
an arbitrary free function; on physical grounds, one can impose some
requirements, via which one can obtain the sought self-force function
uniquely.

To formulate the fundamental condition, let us draw attention to that
the self-force depends on the external force, $\Ls(\Lr,\Lu) = \Ls_\Lf
(\Lr, \Lu)$. It is evident that there is no self-force without external
action; this fact can be taken into account in two ways.

First, we require that if $\Lf$ is zero in a neighborhood of a spacetime
point $\Lrnull$ and four-velocity $\Lrdotnull$ then
 \begin{equation}  \label{qft}
\Ls_\Lf (\Lx_0, \Lu_0) = 0.
 \end{equation}

Second, it is plausible to expect that less action generates less
reaction; consequently, we demand that if the external action tends to
zero then the self-force must tend to zero, too. Specifically, we will
consider the self-force function $\Ls_{\LLlam \Lf}$ for every $0\le
\LLlam \le 1$ and will impose,
in the pointwise sense,
 \begin{equation}  \label{qfr}
\lim_{\LLlam \to 0} \Ls_{\LLlam \Lf} (\Lr, \Lu) = 0.
 \end{equation}
%in the pointwise sense.

A third natural assumption is that if the external field has a
spacetime symmetry then the self-force function has the same symmetry.
Namely, if $\Lf$ is invariant under a Poincar\'e transformation $\LP$
(with the underlying Lorentz transformation $\LL$), i.e.,
%$ \LL^{-1} \Lf (\LP\Lr, \LL \Lu) = \Lf (\Lr, \Lu)$,
$ \LL \Lf
 \!\!\:
\left( \LP^{-1} \Lr, \LL^{-1} \Lu \right) = \Lf (\Lr, \Lu)$,
then the same invariance must hold for $\Ls_\Lf$, too.

Though not utilized in the present considerations, a natural
generalization of this criterion to include noninvariant cases would be
that the transformation of $\Lf$ to
%$\qtraf{\Lf}$,
$\qtraf{\Lf} ( \Lr, \Lu ) = \LL \Lf
% \!\!\;
\left( \LP^{-1} \Lr, \LL^{-1} \Lu \right)$
is accompanied by the corresponding transformation of $\Ls_\Lf$ to
% $\Ls_{\qtraf{\Lf}}$,
$\Ls_{\qtraf{\Lf
 \!\!\;
}} (\Lr, \Lu) = \LL
 \!\;
\Ls_\Lf
 \!\!\:
\left( \LP^{-1} \Lr, \LL^{-1} \Lu \right)$.
 Similarly, our two other conditions could also be generalized/weakened
for future needs.

\subsection{Iteration of the radiation term}  \label{qfb}

Equation \eqref{qen} is, in general, a rather complicated system of
partial differential equations so it is not easy to find its solutions.
Hence, we look for other methods as well, to determine the self-force
function.

An idea is suggested by
% equation
 \eqref{mot1} which, as said, cannot be
an exact equation. We can consider it, however, as a first
approximation. Then it is a straightforward idea that we take an
analogous second approximation: $\dddot \Lr$ is computed as the
derivative of
 \begin{equation}  \label{qfc}
\ddot{\Lr}^j = \LKf^{j} (\Lr, \dot{\Lr}) +
\LLit{\LKs}{1}{j}{} (\Lr, \dot{\Lr}) ,
 \end{equation}
and then $\ddot \Lr$ is substituted by the rhs of \eqref{qfc}. Thus, we
obtain an expression $\Lit{\LKs}{2}(\Lr,\dot \Lr)$, with which the
second approximation for the equation of motion is
 \begin{equation}  \label{qaa}
\ddot{\Lr}^j = \LKf^{j} (\Lr, \dot{\Lr}) +
\LLit{\LKs}{2}{j}{} (\Lr, \dot{\Lr}) .
 \end{equation}
The same procedure can be repeated iteratively for all higher orders. If
the sequence of terms $\Lit{\LKs}{n} (\Lr,\dot \Lr)$ converges to a
$\LKs(\Lr, \dot \Lr)$---in some appropriate sense, e.g.,
% in a
pointwise%
%sense
---then we arrive at a second order equation of motion of the form
\eqref{qav}. Naturally, it is a tough
% mathematical
problem is whether the sequence
% in question
 converges or not.

\subsection{Iteration of the solution}  \label{qfu}

The above iteration method suggests another one, which would not
directly result in an equation of motion but in the motion
corresponding to initial values $\Lrnull$ of spacetime position and
$\Lrdotnull$ of four-velocity.

Let the solution of the zeroth approximation \eqref{qek} corresponding
to initial values $\Lrnull$ and $\Lrdotnull$ be denoted by
$\Lit{\Lz}{0}$. Taking its first and third derivatives, we establish
the differential equation
 \begin{equation}  \label{sec1}
\ddot \Lr^j = \LKf^{j} (\Lr, \dot{\Lr}) + \LKeta \left( {\Lg^j}_k -
\fracc{ \LLit{\dot{\Lz}}{0}{j}{} \LLit{\dot{\Lz}}{0}{\vphantom{j}}{k}
\big}{\Lc^2} \right) \LLit{\dddot{\Lz}\!}{0}{k}{}
 \end{equation}
as the first approximation for the equation of motion. Let
$\Lit{\Lz}{1}$ be its solution for initial values $\Lrnull$ and
$\Lrdotnull$. Taking its first and third derivative, we establish the
second approximation, and so on; at step $n+1$ we solve
 \begin{equation}  \label{secit}
\ddot \Lr^j = \LKf^{j} (\Lr, \dot{\Lr}) + \LKeta \left( {\Lg^j}_k -
\fracc{ \LLit{\dot{\Lz}}{n}{j}{} \LLit{\dot{\Lz}}{n}{\phantom{j}}{k}
}{\Lc^2} \right) \LLit{\dddot{\Lz}\!}{n}{k}{} .
 \end{equation}
If the sequence of solutions $\Lit{\Lz}{n}$ converges to an $\Lz$
then we succeeded in finding a motion satisfying equality \eqref{mot}
without the need for the initial value of acceleration. This motion is,
therefore, a good candidate for the sought physical solution.

Naturally, here, too, convergence is a tough problem.

We can observe that, although this iterative method provides solutions
rather than the equation of motion, a corresponding self-force function
$\Ls (\Lr, \Lu)$ can be read off from the solutions. Namely, the value
of the corresponding $\LKs$ at any spacetime point $\Lrnull$ and
four-velocity value $\Lrdotnull$ can be calculated from the third
derivative of the solution $\Lz$ belonging to initial values $\Lrnull$
and $\Lrdotnull$, at the initial proper time value:
 \begin{equation}  \label{qfa}
\LKs^j (\Lrnull, \Lrdotnull) = \LKeta \left( {\Lg^j}_k -
\fracc{ (\Lrdotnull)^j (\Lrdotnull)_k }{\Lc^2} \right) \dddot{\Lz}^k (0) .
 \end{equation}

 \section{Applying the three approaches in the nonrelativistic regime:
Constant field}  \label{nrit}

Let us consider a constant external electromagnetic field, which acts on
the charged particle via the Lorentz force
 \begin{equation}  \label{zero}
\LLf = \Lq \LLe + \Lq \LLv \times \LLb = \Lq \LLe + \Lq \LLB \LLv ,
 \end{equation}
where $\LLe$ is the electric field three-vector, $\LLb$ is the magnetic
axial vector field, $\LLB = (- \LLb \times)$ is the corresponding
antisymmetric three-tensor---now each assumed space and time
independent---, and $\LLv$ is the velocity of the particle. With the
shorthands
 \begin{equation}  \label{qfd}
\LLHe := \frac{\Lq}{\Lm} \LLe ,
 \qquad
\LLHb := \frac{\Lq}{\Lm} \LLb ,
 \qquad
\LLHB := \frac{\Lq}{\Lm} \LLB ,
 \end{equation}
we can simply write
 \begin{equation}  \label{qfe}
\LLKf = \LLKf (\LLv) = \LLHe + \LLHB \LLv .
 \end{equation}

The electromagnetic nature of the field will not play any role here so
the subsequent considerations will be applicable for any force of the
form \eqref{qfe}, including a constant gravitational attraction, for
example.

We will need some technical remarks regarding $\LLHB$. Its kernel is
spanned by $\LLHb$, its range is the plane orthogonal to $\LLHb$, and,
with $\LLP$ denoting the orthogonal projection onto this plane, we find
 \begin{equation}  \label{qfh}
\LLHB = \LLP \LLHB = \LLHB \LLP = \LLP \LLHB \LLP ,
% \end{equation}
% \begin{equation}  \label{qgd}
 \qquad
(\LLI - \LLP) \LLHB = \LLnull ,
 \qquad
\LLHB^2 = - \LHb^2 \LLP ,
 \end{equation}
where $\LLI$ stands for the three-identity tensor (and $\LHb$ is the
magnitude of $\LLHb$).

To keep the formulae shorter and more easily accessible, we first
treat $\LLHe = \LLnull$. In this case, the equation of motion without
radiation is
 \begin{equation}  \label{qfg}
\dot \LLv =\LLHB \LLv .
 \end{equation}
The results for the general case $\LLHe \not = \LLnull$ are presented in
section~\ref{qfi}.

 \subsection{Differential equation for the self-force function}
\label{qgy}

We can start with ruling out the space and time dependence of $\LLKs$,
based on the requirement of section~\ref{force} that a spacetime
symmetry of the external force should be respected by the self-force,
too. In the present case, the symmetry in question is spacetime
translation invariance. Hence, the sought equation of motion is of the
form
 \begin{equation}  \label{qba}
\dot \LLv = \LLHB \LLv + \LLKs(\LLv) .
 \end{equation}
In addition, the external field is invariant for space inversion,
$-\LLHB(-\LLv)=\LLHB(\LLv)$, and thus $-\LLKs(-\LLv)=\LLKs(\LLv)$ is
required, too.

According to our assumption described in section~\ref{force}, $\LLKs$
must obey the differential equation
 \begin{equation}  \label{qbc}
\LLKs(\LLv) = \LKeta \bigl[ \LLHB + \LLKs'(\LLv) \bigr]
\, \bigl[ \LLHB \LLv + \LLKs(\LLv) \bigr] ,
 \end{equation}
with $'$ denoting the derivative map.
Let us assume that we can expand $\LLKs$ in a series. Because of the
space inversion symmetry, the even powers are zero, so
 \begin{equation}  \label{qeh}
\LLKs (\LLv) = \LLA_1 \LLv + \LLA_3 (\LLv, \LLv,\LLv) + \LLA_5(\LLv,
\LLv, \LLv,\LLv,\LLv) + \cdots
 \end{equation}
where $\LLA_1$ is a linear map, $\LLA_3$ is a symmetric trilinear map
etc.; keep in mind that they depend on $\LLHB$. Using the notation
$\LLA_3 \big( \LLv^3 \big) := \LLA_3 (\LLv, \LLv,\LLv)$ etc., we obtain
 \begin{align}  \label{qbd}
\LLA_1 \LLv + \LLA_3 \big( \LLv^3 \big) + \LLA_5 \big( \LLv^5 \big) + \cdots
 &
=
\LKeta \bigl[ \LLHB + \LLA_1 + 3 \LLA_3 \big( \LLv^2, \,\cdot\, \big) + 5
\LLA_5 \big( \LLv^4, \,\cdot\, \big) + \cdots \bigr]
 \nn \\
 &
\quad
 \times
\bigl[ \LLHB \LLv + \LLA_1 \LLv + \LLA_3 \big( \LLv^3 \big) + \LLA_5
\big( \LLv^5 \big) + \cdots \bigr],
 \hskip 1.7em
 \end{align}
from which it follows, order by order, that
 \begin{align}  \label{A1}
\LLA_1
 &
=
\LKeta \big( \LLHB + \LLA_1 \big)^2,
 \\   \label{A3}
\LLA_3 \big( \LLv^3 \big)
 &
=
\LKeta \bigl[ 3\LLA_3 \big( \LLv^2, \big( \LLHB + \LLA_1 \big) \LLv \big)
+ \big( \LLHB + \LLA_1 \big) \LLA_3 \big( \LLv^3 \big) \bigr],
 \\   \label{A5}
\LLA_5 \big( \LLv^5 \big)
 &
=
\LKeta \bigl[ 3 \LLA_3 \big( \LLv^2, \LLA_3 \big( \LLv^2 \big) \big) +
5 \LLA_5 \big(\LLv^4, \big( \LLHB + \LLA_1 \big) \LLv \big)
% \nn\\   \label{qge}
% &
% \quad
+ \big( \LLHB + \LLA_1 \big) \LLA_5 \big( \LLv^5 \big) \bigr],
 \end{align}
etc. Multiplying \eqref{A1} from the left and from the right by $\LLHB +
\LLA_1$, we find that $\LLA_1 \LLHB = \LLHB \LLA_1$. Then it is a simple
algebraic fact that
 \begin{equation}  \label{qbe}
\LLA_1 = - \Lbet \LLHB - \Lalp \LLP + \LLLlam (\LLI - \LLP),
 \end{equation}
where $\Lbet$, $\Lalp$ and $\LLLlam$ are scalar coefficients depending
on $\LLHB$. Applying the second condition put in section~\ref{force},
$\Lalp$ and $\LLLlam$ must tend to zero if $\LLHB$ tends to zero.

Having \eqref{qbe}, \eqref{A1} yields
 \begin{align}  \label{qbf}
- \Lbet \LLHB - \Lalp
\LLP + \LLLlam (\LLI - \LLP)
% &
% \nn\\   \label{qgf}
% &
=
\LKeta \bigl[ - (1 - \Lbet)^2 \LHb^2 \LLP + \Lalp^2 \LLP +\LLLlam^2
(\LLI - \LLP) - 2 \Lalp (1 - \Lbet) \LLHB \bigr] ,
 \end{align}
which tells
 \begin{equation}  \label{huha}
\Lbet = 2 \LKeta \Lalp (1 - \Lbet) ,
 \qquad
\Lalp = \LKeta (1 - \Lbet)^2 \LHb^2 - \LKeta\Lalp^2
 \end{equation}
as well as $\LLLlam = \LKeta \LLLlam^2$, according to which $\LLLlam$ is
either zero or equals $\fracc1{\LKeta}$. The second possibility is
excluded by the
%fact
% requirement
demand
that $\LLLlam$ must be zero for zero external field so
 \begin{equation}  \label{qbg}
\LLLlam = 0.
 \end{equation}
The first equation in \eqref{huha} rules out $\Lbet = 1$, and then
another equivalent pair of equations is
 \begin{equation}  \label{gamma}
\LKeta \Lalp
=
\frac{\Lbet}{2(1 - \Lbet)} ,
% \end{equation}
% \begin{equation}  \label{qgg}
\qquad
4 (\LKeta \LHb)^2 (1 - \Lbet)^4 + (1 - \Lbet)^2 - 1
=
0 .
 \end{equation}
The latter condition is a quadratic equation for $\Lphi := (1 - \Lbet)^2
\ge 0$, with the only non-negative root
 \begin{equation}  \label{qcu}
%\Lphi = \frac{-1 + \sqrt{1 + 16(\LKeta \LHb)^2}}{8(\LKeta \LHb)^2} .
\Lphi = \left( -1 + \sqrt{ 1 + \hbox{\raisebox{0ex}[1.7ex]
{$16\big(\LKeta \LHb\big)^2$}} } \hskip .3em \right) \Big/
\left( 8 \big(\LKeta \LHb\big)^2 \right) .
 \end{equation}
%From here, we derive $ \Lbet = 1 \pm \sqrt{ \Lphi } $.

Again, the condition that $\Lalp$ must be zero for zero external field,
gives the result:
 \begin{equation}  \label{betalp}
\Lbet = 1 - \sqrt{\Lphi} ,
 \qquad
%\Lalp = \fraccc{1}{2\LKeta} \left( \fracc{1}{ \sqrt{\Lphi}} - 1 \right) ,
\Lalp = \fraccc{ \left( \fracc{1}{ \sqrt{\Lphi}} - 1 \right) }{2\LKeta} ,
 \end{equation}
the latter obtained from the first equation in \eqref{gamma}.
As it is proved in the Appendix, \eqref{A3} yields $\LLA_3 = \LLnull$,
and, similarly, all the higher order terms are found to be zero. Hence,
the self-force is
 \begin{equation}  \label{qep}
\LLKs (\LLv) = -(\Lbet \LLHB + \Lalp \LLP)\LLv,
 \end{equation}
and the equation of motion is
 \begin{equation}  \label{joegy}
\dot \LLv = \LLHB \LLv + \LLKs (\LLv),
% \end{equation}
 \quad \hbox{i.e.,}  \quad
%i.e.,
% \begin{equation}  \label{qgw}
\dot \LLv = \bigl[ (1 - \Lbet) \LLHB - \Lalp \LLP \bigr] \LLv .
 \end{equation}

It is informative to inspect the solution of this equation of motion,
which is
 \begin{equation}  \label{prp}
\LLw(\Lt) = (\LLI - \LLP) \LLc + \e^{-\Lalp \Lt} \e^{ (1 - \Lbet)
\LLHB \Lt } \LLP \LLc
 \end{equation}
for initial velocity $\LLc$ at zero initial time. For nonzero $\LHb$, we
have $0 < 1 - \Lbet < 1$ and $\Lalp > 0$ so radiation causes that
 \begin{itemize}
 \item
the effect of the external magnetic field is reduced by a certain
factor, and
 \item
the component of velocity perpendicular to the magnetic field tends to
zero as time passes.
 \end{itemize}

Note that, in this simple case, two conditions given at the end of
section~\ref{force} suffice to determine the self-force function
completely.

\subsection{Iteration of the radiation term}

It follows from the equation without radiation---the zeroth
approximation \eqref{qfg}---that $\ddot \LLv = \LLHB \dot \LLv = \LLHB^2
\LLv=- \LHb^2 \LLP\LLv $. Accordingly, the first approximation of the
radiation term is
 \begin{equation}  \label{qdc}
\Lit{\LLKs}{1}(\LLv) = \Lit{\LLK}{1} \LLv ,
 \qquad
\Lit{\LLK}{1} = - \LKeta \LHb^2 \LLP ,
 \end{equation}
with which the first approximation of the equation of motion becomes
$\dot \LLv = \bigl( \LLHB + \Lit{\LLK}{1} \bigr) \LLv$. The
radiation term derived from this equation equals
 \begin{equation}  \label{qfj}
\Lit{\LLKs}{2}(\LLv) = \Lit{\LLK}{2} \LLv ,
 \qquad
\Lit{\LLK}{2} = \LKeta \bigl( \LLHB + \Lit{\LLK}{1} \bigr)^2 .
 \end{equation}
Repeating this again and again, we recognize the generic recursion
formula
 \begin{equation}  \label{qfk}
\Lit{\LLKs}{n}(\LLv) = \Lit{\LLK}{n} \LLv ,
 \qquad
\Lit{\LLK}{n} = \LKeta \bigl( \LLHB + \Lit{\LLK}{n-1} \bigr)^2 .
 \end{equation}
Since
 \begin{equation}  \label{qdd}
\Lit{\LLK}{2} = \LKeta \bigl( \LLHB + \Lit{\LLK}{1} \bigr)^2 =
\LKeta \bigl( - \LHb^2 \LLP - 2 \LKeta \LHb^2 \LLHB + \LKeta^2 \LHb^4
\LLP \bigr) ,
 \end{equation}
every $\Lit{\LLK}{n}$ is a linear combination of $\LLHB$ and
$\LLP$.
Supposing that the sequence $\Lit{\LLKs}{n}(\LLv)$ converges for all
$\LLv$,  $\Lit{\LLK}{n}$ should converge to an $\LLK$, for which we have
 \begin{equation}  \label{qcw}
\LLK = \LKeta(\LLHB + \LLK)^2 ,
 \end{equation}
where $\LLK$ is a linear combination of $\LLHB$ and $\LLP$:
% \begin{equation}  \label{qcx}
$
\LLK = - \Lbet \LLHB - \Lalp \LLP.
$
% \end{equation}
Consequently,
 \begin{equation}  \label{qcy}
- \Lbet \LLHB - \Lalp \LLP = \LKeta \left[ - (1 - \Lbet)^2 \LHb^2 \LLP -
2 \Lalp (1 - \Lbet) \LLHB + \Lalp^2 \LLP\right] ,
 \end{equation}
which imposes the pair of equations in \eqref{huha}, i.e. we arrive at
the same result as previously.

The ambiguity $ \Lbet = 1 \pm \sqrt{ \Lphi } $ arises here, too. If
convergence holds then, naturally, only one of the possibilities can be
the limit.  To select the correct one, we can consider the simple case
when there is no external force.  Then all the iteration terms are zero,
and the iteration converges trivially. Hence, the coefficients must be
zero for zero magnetic field, as previously.

A problem with the method of iterating the radiation term is that it is
%difficult
hard to obtain conditions for the convergence of the iteration.
Unfortunately, convergence
% does not hold necessarily.
may not hold. Indeed, for
$\LKeta \LHb = 1$---i.e., for such special magnetic fields---convergence
does not occur, since then $\Lit{\LLK}{2} = -2\LLHB $, so $\Lit{\LLK}{3}
= \Lit{\LLK}{1}$. %Consequently,
Thus, for all higher $n$s,
 \begin{equation}  \label{qeo}
\Lit{\LLK}{n} =
 \begin{cases}
\Lit{\LLK}{2}
 \quad
(n \ \hbox{is even}),
 \\
\Lit{\LLK}{1}
 \quad
(n \ \hbox{is odd}).
 \end{cases}
 \end{equation}
It is interesting, however, that the final result \eqref{betalp}  does
not exclude the case $\LKeta \LHb = 1$, and provides solution for
$\LKeta \LHb > 1$ as well (which is expected to be outside the domain of
convergence).

\subsection{Iteration of the solution}

Considering an initial value $\LLc$ at zero initial time, the zeroth
equation \eqref{qfg} has the solution
 \begin{equation}  \label{qch}
\Lit{\LLw}{0} (\Lt) = \e^{\LLHB \Lt} \LLc.
 \end{equation}
Then $\Lit{\ddot\LLw}{0} (\Lt) = -\LHb^2 \e^{\LLHB \Lt} \LLc$, so the
first approximation satisfies the equation
 \begin{equation}  \label{qci}
\dot \LLv = \LLHB \LLv - \LKeta \LHb^2 \e^{\LLHB \Lt} \LLc ,
 \end{equation}
whose solution with initial value $\LLc$ is
 \begin{equation}  \label{qcj}
\Lit{\LLw}{1} (\Lt) = \left( 1 - \LKeta \LHb^2 \Lt \right) \e^{\LLHB \Lt} \LLc.
 \end{equation}
Then $ \Lit{\ddot \LLw}{1} (\Lt) = - \left[ \LHb^2 \left( 1 - \LKeta
\LHb^2 \Lt \right) + 2\LKeta \LHb^2 \LLHB \right] \e^{\LLHB \Lt} \LLc$.
The second approximation satisfies the equation
 \begin{equation}  \label{qck}
\dot \LLv = \LLHB \LLv - \LKeta \left[ \LHb^2 \big( 1 - \LKeta \LHb^2\Lt
\big) + 2 \LKeta \LHb^2 \LLHB \right] \e^{\LLHB \Lt} \LLc ,
 \end{equation}
with the solution
 \begin{equation}  \label{qcl}
\Lit{\LLw}{2} (\Lt) = \left[ 1 - \LKeta \LHb^2 \left( \Lt - \frac{1}{2}
\LKeta \LHb^2 \Lt \right) - 2 \LKeta^2 \LHb^2 \Lt \, \LLHB \right]
\e^{\LLHB \Lt} \LLc .
 \end{equation}
At the $n$th step of this iteration scheme, we find
 \begin{equation}  \label{qcm}
\Lit{\LLw}{n} (\Lt) = \left[ \Lit{\LLp}{n} (\Lt) +
\Lit{\LLq}{n} (\Lt) \LLHB \right] \e^{\LLHB \Lt} \LLc ,
 \end{equation}
where $\Lit{\LLp}{n}$ and $\Lit{\LLq}{n}$ are polynomials of
$\Lt$ satisfying the recursive formulae
 \begin{align}  \label{qcn}
\Lit{\dot \LLp}{n}
 &
=
\LKeta \left( \Lit{\ddot \LLp}{n-1} - 2
\LHb^2 \Lit{\dot \LLq}{n-1} - \LHb^2 \LLp_{n-1} \right) ,
 \\   \label{qco}
\Lit{\dot \LLq}{n}
 &
=
\LKeta \left( \Lit{\ddot \LLq}{n-1} + 2
\Lit{\dot \LLp}{n-1} - \LHb^2 \Lit{\LLq}{n-1} \right)
 \end{align}
and the initial conditions
$\Lit{\LLp}{n} (0) = 1, \; \Lit{\LLq}{n} (0) = 0$.

Let us suppose that $\LLLf := \lim_n \Lit{\LLp}{n}$ and $\LLLg :=
\lim_n \Lit{\LLq}{n}$ exist and, moreover, that the limit procedure
and differentiation can be interchanged.
%Then we have for $\LLw := \lim_n \Lit{\LLw}{n}$
Then, for $\LLw := \lim_n \Lit{\LLw}{n}$,
 \begin{align}  \label{sec}
\LLw(\Lt)
 &
=
\bigl[ \LLLf(\Lt) + \LLLg(\Lt) \LLHB \bigr] \e^{\LLHB \Lt} \LLc ,
 \\   \label{qcp}
\dot \LLLf
 &
=
\LKeta \left( \ddot \LLLf - 2\LHb^2 \dot \LLLg - \LHb^2 \LLLf \right),
 \\   \label{qgh}
\dot \LLLg
 &
=
\LKeta \left( \ddot \LLLg + 2\dot \LLLf - \LHb^2 \LLLg \right)
 \end{align}
together with the conditions $\LLLf (0) = 1, \; \LLLg (0) = 0$. With the
notation $ \LLLs := \LLLf + \ii \LHb \LLLg $, \eqref{qcp} and
\eqref{qgh} can be comprised as
 \begin{equation}  \label{qfo}
\ddot \LLLs - \left(\fracc{1}{\LKeta} - 2\ii \LHb \right) \dot \LLLs -
\LHb^2 \LLLs = 0 .
 \end{equation}
The roots of the corresponding characteristic polynomial are
 \begin{equation}  \label{qfp}
\fracc{1}{2\LKeta} - \ii \LHb \pm \sqrt{ \fraccc{1}{4\LKeta^2} -
\ii \fracc{\LHb}{\LKeta} }.
 \end{equation}

The emerging ambiguity can be resolved as previously: if the magnetic
force is sent to zero then $\Lit{\LLp}{n}(\Lt) = 1$ for all $n$,
therefore, we choose the root that vanishes for $\LHb \to 0$ so as to
obtain the solution $\LLLf (\Lt) = 1$, i.e., the free motion (no
self-force).

Therefore, if $(-\Lalp)$ and $\Lbet$ are the real and the imaginary part
of the root, respectively, then the solution of our system of
differential equations is
 \begin{equation}  \label{qcr}
\LLLf (\Lt) = \e^{- \Lalp \Lt} \cos(\Lbet \LHb \Lt),
 \qquad
\LLLg (\Lt) = - \e^{- \Lalp \Lt} \fracc{\sin(\Lbet \LHb \Lt)}{\LHb},
 \end{equation}
where $\Lbet$ and $\Lalp$ satisfy the relations in \eqref{huha}.

We have obtained the same result as previously. In particular, utilizing
\eqref{qfh}%
%--\eqref{qgd}
, one can demonstrate that \eqref{sec} is the same as \eqref{prp}.
At last, it is straightforward to find that the self-force function
corresponding to \eqref{sec}---see \eqref{qfa}---proves to be the
same as \eqref{qep}.

On the other side, one can also observe that, though leading to the
same result, the two iterations themselves are different: the
solutions of the iterated equations $\dot \LLv = \bigl( \LLHB +
\Lit{\LLK}{n} \bigr) \LLv$ do not equal the functions \eqref{qcm}.

%Naturally,
Convergence is a nontrivial question in this approach, too.
Nevertheless, it is interesting that, with this method, there is no
evidence for excluding the case $ \LKeta \LHb = 1 $.

\subsection{Nonzero electric and magnetic field}  \label{qfi}

As anticipated in section~\ref{nrit}, now we turn towards the case of a
nonzero constant electric field in addition to the nonzero constant
magnetic field.

It is beneficial to decompose the electric field into components
parallel to and orthogonal to the magnetic field, respectively, and to
observe that this decomposition can be written in the form
 \begin{equation}  \label{qfw}
\LLHe = (\LLI - \LLP) \LLHe + \LLHB \left( - \frac{1}{\LHb^2} \LLHB \LLHe
\right) .
 \end{equation}
This induces a decomposition of the equation of motion without
self-force [i.e., containing only the external Lorentz force
\eqref{qfe}]. The $\LLHb$-parallel component,
 \begin{equation}  \label{qfx}
[ (\LLI - \LLP) \LLv ] \dot{\phantom{i}} = (\LLI - \LLP) \LLHe ,
 \end{equation}
governs only the $\LLHb$-parallel component of $\LLv$, while the
$\LLHb$-orthogonal part can be written as
 \begin{equation}  \label{qfy}
\dot\LLu = \LLHB \LLu
% \end{equation}
 \qquad \hbox{with} \qquad
%with
% \begin{equation}  \label{qgi}
\LLu := \LLP \LLv - \frac{1}{\LHb^2} \LLHB \LLHe ,
 \qquad
\LLu \perp \LLHb ,
 \end{equation}
and determines the time evolution of the $\LLHb$-orthogonal component of
$\LLv$.

Now we add the self-force term. Both iteration methods, namely, that
of the radiation term and that of the solution, can be found to preserve
this decomposition, where the $\LLHb$-orthogonal part can actually be
treated the same way for $\LLu$ as we proceeded in the $\LLHe = \LLnull$
case for $\LLv$. Both approaches provide
 \begin{equation}  \label{qfz}
\LLKs (\LLv) = - \Lbet \LLP \LLHe + \Lalp \frac{1}{\LHb^2} \LLHB \LLHe -
\big[ \Lbet \LLHB + \Lalp \LLP \big] \LLv
 \end{equation}
for the self-force and
 \begin{align}  \label{qga}
\LLv(\Lt)
% &
=
(\LLI - \LLP) \LLHe \Lt + (\LLI - \LLP) \LLc + \frac{1}{\LHb^2} \LLHB \LLHe
% \nn\\   \label{qgj}
% &
% \quad
+ \e^{-\Lalp \Lt} \e^{ (1 - \Lbet) \LLHB \Lt } \left( \LLP \LLc -
\frac{1}{\LHb^2} \LLHB \LLHe \right)
 \end{align}
for the solution, after putting the two decomposed parts together. The
found self-force function proves to be a solution of the partial
differential equation of the third approach as well, and satisfies the
two additional requirements---vanishing for vanishing external field
and symmetry preservation.

\section{Applying the three approaches in the nonrelativistic regime:
Harmonic force}  \label{qeg}

As our second physical system considered as example for the three
proposed methods, we next investigate the one-dimensional nonrelativistic
motion due to a harmonic elastic force. Without radiation back-reaction,
the equation is
 \begin{equation}  \label{qdh}
\ddot \Lx = -\Lome^2 \Lx ,
 \end{equation}
where $\Lome$ is a non-negative constant.

\subsection{Differential equation for the self-force function}

According to our assumption described in section~\ref{force}, the
radiation self-force function $\LKs(\Lx, \dot \Lx)$ in the anticipated
equation of motion
 \begin{equation}  \label{qdp}
\ddot \Lx = - \Lome^2 \Lx + \LKs(\Lx, \dot \Lx)
 \end{equation}
is to satisfy the quasi-linear partial differential equation
 \begin{equation}  \label{qdq}
\LKs (\Lx, \Lv) = \LKeta \left[ - \Lome^2 \Lv + \frac{\partial \LKs
(\Lx, \Lv)}{\partial \Lx} \Lv + \frac{\partial \LKs (\Lx, \Lv)}{\partial
\Lv} \left( - \Lome^2 \Lx + \LKs \right) \right] .
 \end{equation}
Note that, although not denoted explicitly, the sought $\LKs$ also
depends on $\Lome$, i.e., on the external force.

The characteristic ordinary differential equation corresponding to
\eqref{qdq} reads
 \begin{equation}  \label{char}
\frac{\dd \Lx}{\dd \Lxi} = \Lv,
 \qquad
\frac{\dd \Lv}{\dd \Lxi} = - \Lome^2 \Lx + \LKs,
 \qquad
\frac{\dd \LKs}{\dd \Lxi} = \frac{1}{\LKeta} \LKs + \Lome^2 \Lv .
 \end{equation}
This is a simple linear differential equation whose characteristic roots
$\Llam$ fulfill the equation
 \begin{equation}  \label{lam}
\LKeta \Llam^3 - \Llam^2 - \Lome^2 = 0.
 \end{equation}
We can find its solutions (roots) by the Cardano formula.
With the notation
 \begin{equation}  \label{qer}
\Lro_{\pm} :=
{\rule{0em}{3.8ex}}^{3} \hskip-.48em  %% \sqrt[3]  looks ugly !!!
\sqrt{ \fracc{\LKeta^2\Lome^2}{2} + \fracc{1}{27}
\pm \sqrt{ \fracc{\LKeta^4\Lome^4}{4} + \fracc{\LKeta^2\Lome^2}{27}}} ,
 \end{equation}
the roots are
 \begin{align}  \label{qes}
\LKeta\Llam_1
 &
=
% &
\frac1{3} - \frac{\Lro_+ + \Lro_-}{2} + \ii \frac{\sqrt{3}
(\Lro_+ - \Lro_-)}{2},
 \\   \label{qgk}
\LKeta\Llam_2
 &
=
% &
\frac1{3} - \frac{\Lro_+ + \Lro_-}{2} - \ii \frac{\sqrt{3}
(\Lro_+ - \Lro_-)}{2} ,
 \\   \label{qet}
\LKeta\Llam_3
 &
=
% &
\frac1{3} + \Lro_+ + \Lro_- .
 \end{align}
With the three roots, the solutions of equation \eqref{char} are of the
form
 \begin{align}  \label{qeu}
\Lx (\Lxi)
 &
=
% &
\sum_{i=1}^3 \LLLLa_i \e^{\Llam_i \Lxi} ,
% \\   \label{qgm}
&
\Lv (\Lxi)
 &
=
% &
\sum_{i=1}^3 \Llam_i \LLLLa_i \e^{\Llam_i \Lxi} ,
% \\   \label{qgl}
&
\LKs (\Lxi)
 &
=
% &
\sum_{i=1}^3 \Llam_i^2 \LLLLa_i \e^{\Llam_i \Lxi} + \Lome^2 \Lx (\Lxi) .
 \end{align}
According to the method of characteristics, via eliminating the
auxiliary variable $\Lxi$, one obtains $\LKs$ as a function of $\Lx$ and
$\Lv$. Now we use the condition of section~\ref{force} that $\LKs$ must be
zero for zero external force i.e. for $\Lome = 0$. Since $\Llam_1$ and
$\Llam_2$  are zero for $\Lome = 0$ and $\Llam_3$ is not zero, we deduce
from the last equality above that $\LLLLa_3 = 0$ is necessary.

Then $\Lxi$ can be eliminated easily; with the notations $\LLLLb_1 :=
\LLLLa_1 \e^{\Llam_1 \Lxi}$ and $\LLLLb_2 := \LLLLa_2 \e^{\Llam_2
\Lxi}$, we have
 \begin{align}  \label{qev}
\Lx
 &
=
% &
\LLLLb_1 + \LLLLb_2 ,
% \\   \label{qgn}
&
\Lv
 &
=
% &
\Llam_1 \LLLLb_1 + \Llam_2 \LLLLb_2 ,
% \\   \label{qgo}
&
\LKs
 &
=
% &
\Llam_1^2 \LLLLb_1 + \Llam_2^2 \LLLLb_2 + \Lome^2 \Lx .
 \end{align}
Here, the first two equations enable one to express $\LLLLb_1$ and
$\LLLLb_2$ as a linear function of $\Lx$ and $\Lv$. Substituting them
into the third equation provides $\LKs$ also as a linear function of
$\Lx$ and $\Lv$: \ $\LKs(\Lx, \Lv) = (\Lome^2 - \Llam_1\Llam_2) \Lx +
(\Llam_1 + \Llam_2) \Lv$.

A convenient way to proceed is to write the coefficients in another
form:
 \begin{equation}  \label{qew}
\LKs(\Lx, \Lv) = \Lbet \Lome^2 \Lx - \Lalp \Lv .
 \end{equation}
Evidently,
 \begin{equation}  \label{qal}
\LKeta\Lalp = -\frac2{3} + \Lro_+ + \Lro_- ,
 \end{equation}
and, substituting \eqref{qew} into \eqref{qdq}, we obtain
 \begin{equation}  \label{ezaz}
\Lbet = \frac{\LKeta \Lalp}{1 + \LKeta \Lalp} \qquad \hbox{and} \qquad
\LKeta \Lalp (1 + \LKeta \Lalp)^2 = \LKeta^2 \Lome^2.
 \end{equation}
 We can see that $\Lalp > 0$ and $0 < \Lbet < 1$ are necessary for $\Lome
\neq 0$.
Hence, the equation of motion with the radiation term \eqref{qew} reads
 \begin{equation}  \label{joegye}
\ddot \Lx = - (1 - \Lbet) \Lome^2 \Lx - \Lalp \dot \Lx .
 \end{equation}
According to this equation, radiation causes that
 \begin{itemize}
 \item
the effect of the harmonic force is reduced by a certain factor, and
 \item
the motion is damped by a term proportional to velocity.
 \end{itemize}

We can see in this case, too, that the condition that the
self-force must be zero if the external force is zero suffices to
determine the self-force function completely.

\subsection{Iteration of the radiation term}  \label{qfm}

It follows from the equation without radiation (the zeroth
approximation) \eqref{qdh} that $\dddot \Lx =- \Lome^2 \dot \Lx$, so the
the first  approximation of the radiation term is
 \begin{equation}  \label{qdce}
\Lit{\LKs}{1} (\Lx, \dot \Lx) = -\LKeta \Lome^2 \dot \Lx,
 \end{equation}
and the first approximate equation of motion becomes $\ddot \Lx =
-\Lome^2 \Lx + \Lit{\LKs}{1} (\Lx, \dot \Lx)$. Computing $\dddot \Lx$
from this  equation and then replacing $\ddot \Lx$ with $- \Lome^2 \Lx +
\Lit{\LKs}{1} (\Lx, \dot \Lx)$, we obtain the second approximation
 \begin{equation}  \label{qdk}
\Lit{\LKs}{2} (\Lx, \dot \Lx) = \LKeta^2 \Lome^4 \Lx - \LKeta \Lome^2
\left( 1 - \LKeta^2 \Lome^2 \right) \dot \Lx
 \end{equation}
and the second equation $\ddot \Lx = -\Lome^2 \Lx + \Lit{\LKs}{2} (\Lx,
\dot \Lx)$.

It is straightforward then that the $n$th approximation is of the form
 \begin{equation}  \label{qdl}
\Lit{\LKs}{n} (\Lx,\dot \Lx) = \Lbet_n \Lome^2 \Lx - \Lalp_n \dot \Lx ,
 \end{equation}
where $\Lbet_n$ and $\LKeta \Lalp_n$ are functions (polynomials) of
$\LKeta^2 \Lome^2$ and satisfy the recursive formulae
 \begin{align}  \label{qdm}
\Lbet_{n+1}
 &
=
% &
\LKeta \Lalp_n (1 - \Lbet_n) ,
% \\   \label{qgp}
&
\LKeta \Lalp_{n+1}
 &
=
% &
\LKeta^2 \Lome^2 (1 - \Lbet_n) - (\LKeta \Lalp_n)^2.
 \end{align}
Supposing that the sequence $\Lit{\LKs}{n}$ converges to a $\LKs$,
i.e., the limits $\Lbet := \lim_n \Lbet_n$ and $\Lalp := \lim_n \Lalp_n$
exist, then
 \begin{equation}  \label{lims}
\LKs(\Lx, \dot \Lx) = \Lbet \Lome^2 \Lx - \Lalp \dot \Lx
 \end{equation}
and
 \begin{equation}  \label{ezisaz}
\Lbet = \frac{\LKeta \Lalp}{1 + \LKeta \Lalp},
 \qquad
\LKeta \Lalp (1 + \LKeta \Lalp)^2 = \LKeta^2 \Lome^2 ,
 \end{equation}
which coincide with \eqref{ezaz}. Putting $\LKeta\Llam:=1+ \LKeta\Lalp$,
the second equation above is transformed into the equation \eqref{lam}.
Naturally, now we are interested in the real roots. We find that only
one root is real, namely, with the notation \eqref{qer},
 \begin{equation}  \label{qfss}
\LKeta \Lalp = -\frac2{3} + \Lro_+ + \Lro_- .
 \end{equation}
Hence, we arrive at the same result as previously.

It is a problem, though, that it is difficult to obtain conditions for
the convergence of the iteration. Unfortunately, convergence does not
hold necessarily. Indeed, if $\LKeta \Lome = 1$ then the sequence does
not converge because then $\Lit{\LKs}{2} (\Lx, \dot \Lx) = \Lome \Lx^2$,
and thus $\Lit{\LKs}{3} = \Lit{\LKs}{1}$. Consequently,
 \begin{equation}  \label{qeq}
\Lit{\LKs}{n} =
\begin{cases}
 \Lit{\LKs}{2} \quad (n \ \hbox{is even}),
\\
 \Lit{\LKs}{1} \quad (n \ \hbox{is odd}).
\end{cases}
 \end{equation}
It is interesting, however, that, with \eqref{qfss}, \eqref{ezisaz}
provides a solution for all values of $\LKeta\Lome$.

\subsection{Iteration of the solution}

Considering some initial values $\LLLa$ and $\LLLh$ for position and
velocity, the zeroth equation
\eqref{qdh} has the solution
 \begin{equation}  \label{qel}
\Lit{\Lz}{0} (\Lt) = \LLLc \e^{\ii \Lome \Lt} + \LLLc^* \e^{-\ii \Lome
\Lt} = \LLLc \e^{\ii \Lome \Lt} + \cc
% \qquad  \hbox{with}  \quad
 \end{equation}
with
 \begin{equation}  \label{qgx}
%\LLLc := \frac{1}{2} \left(\LLLa - \frac{\ii \LLLh}{\Lome} \right) .
\LLLc := \left( \LLLa - \fracc{\ii \LLLh}{\Lome} \right) / 2.
 \end{equation}
Since $\Lit{\dddot \Lz}{0} (\Lt) = -\ii \LLLc \Lome^3 \e^{\ii \Lome
\Lt} + \cc$, the equation for the first iterated solution reads
 \begin{equation}  \label{qdt}
\ddot \Lx = -\Lome^2 \Lx + \LKeta \left( - \ii \LLLc \Lome^3 \e^{\ii
\Lome \Lt} + \cc \right) ,
 \end{equation}
whose solution with the chosen initial values is
 \begin{equation}  \label{qdtt}
\Lit{\Lz}{1} (\Lt) =
\left[ -({1}/{2}) \LKeta \Lome^3 \Lt + \LLLc \right] \e^{\ii \Lome
\Lt} + \cc
 \end{equation}
Then we find that the $n$th solution is of the form
 \begin{equation}  \label{qdu}
\Lit{\Lz}{n} (\Lt) = \LLp_n (\Lt) \e^{\ii \Lome \Lt} + \cc ,
 \end{equation}
where $\LLp_n (\Lt)$ is a polynomial of $\Lt$ of $n$th degree; and we
have
 \begin{equation}  \label{qdv}
\Lit{\ddot \Lz}{n} = ( \ddot \LLp_n + 2 \ii \Lome \dot \LLp_n -
\Lome^2 \LLp_n ) \e^{\ii \Lome \Lt} + \cc
 \end{equation}
and
 \begin{equation}  \label{qdw}
\Lit{\dddot \Lz}{n} = ( \dddot \LLp_n + 3 \ii \Lome \ddot \LLp_n - 3
\Lome^2 \dot \LLp_n - \ii \Lome^3 \LLp_n ) \e^{\ii \Lome \Lt} + \cc \,,
 \end{equation}
and the resulting recursive formula (arising both from the coefficient
of $\e^{\ii \Lome \Lt}$ and from that of $\e^{- \ii \Lome \Lt}$) is
 \begin{align}  \label{qdx}
\ddot \LLp_{n+1} + 2 \ii \Lome \dot \LLp_{n+1} - \Lome^2 \LLp_{n+1}
% \nn\\   \label{qgr}
= -
\Lome^2 \LLp_{n+1} + \LKeta ( \dddot \LLp_n + 3 \ii \Lome \ddot \LLp_n -
3 \Lome^2 \dot \LLp_n - \ii \Lome^3 \LLp_n ) .
 \end{align}
Thus, supposing convergence (and that differentiation can be
interchanged with taking the limit), we find for the limit $ \Lp :=
\lim_n \LLp_n $
 \begin{align}  \label{qdy}
\ddot \Lp + 2 \ii \Lome \dot \Lp - \Lome^2 \Lp
% \nn\\   \label{qgs}
= - \Lome^2 \Lp + \LKeta ( \dddot \Lp + 3 \ii \Lome \ddot \Lp - 3
\Lome^2 \dot \Lp - \ii \Lome^3 \Lp ) .
 \end{align}
This is a linear differential equation of third order. Its solutions are
of the form $\e^{\Llam t}$ with
 \begin{equation}  \label{qdz}
( \Llam + \ii \Lome )^2 + \Lome^2 = \LKeta ( \Llam + \ii \Lome )^3.
 \end{equation}
Putting $\Llam + \ii \Lome = \Lmu + \ii \Lnu$, where $\Lmu$ and $\Lnu$
are real, we can rewrite this as
 \begin{equation}  \label{qea}
\Lmu^2 + 2 \ii \Lmu \Lnu - \Lnu^2 + \Lome^2 = \LKeta ( \Lmu^3 + 3 \ii
\Lmu^2 \Lnu - 3 \Lmu \Lnu^2 - \ii \Lnu^3 ),
 \end{equation}
which is equivalent to the pair of real equations
 \begin{gather}  \label{qeb}
\Lnu^2 = 3\Lmu^2 - \fracc{2\Lmu}{\LKeta} ,
% \end{align}
% \begin{align}
 \\  \label{qgt}
8 ( \LKeta \Lmu )^3 - 8 ( \LKeta \Lmu )^2 + 2 ( \LKeta \Lmu ) - ( \LKeta
\Lome )^2 = 0 .
 \end{gather}
With $\Lalp := 2 \Lmu$, \eqref{qgt} reduces to the second equation of
\eqref{ezisaz}. In parallel, putting $ \Lnu^2 =: (1 - \Lbet) \Lome^2 -
{\Lalp^2}/{4} $ (i.e., defining $\Lbet$ in this way), we find $ \LKeta
\Lalp (\LKeta \Lalp + 1) = (1 - \Lbet) \Lome^2 $, which, together with
the second equation of \eqref{ezisaz}, results in its first one.

As a consequence, whenever convergence holds, the iteration of solutions
gives the same solutions and same self-force function as the iteration
of the radiation term.

\section{Discussion}

We looked for a second order equation of motion whose solutions satisfy
the Lorentz--Dirac equality and, at the same time, are physically
acceptable. Our aim was to give back-reaction as a function of spacetime
position and velocity. A simple argument showed that this self-force
function is determined by a first order partial differential equation.
Two iterative methods, too, were proposed for finding the self-force
function.

In the nonrelativistic approximation we could exactly calculate the
self-force function for two systems: a constant external electromagnetic
field and a one-dimensional elastic external force. The three suggested
methods turned out to lead to the same result. As concerns the physical
picture, for both systems, radiation back-reaction has two
manifestations: inducing a damping linear in velocity and reducing the
strength of the external force.

The latter effect could also allow the---quantum field theory
motivated---interpretation that back-reaction causes a positive
renormalization of the mass of the particle. However, for
the constant external electromagnetic field, this renormalization
turns out to differ from a simple scalar multiplying of the mass.
Rather, renormalization is a direction dependent, tensorial
multiplication.

This system also proves to show a limitation of the criterion by Dirac
and Haag \cite{Dirac,Haag}, which would choose that solution for initial
position and velocity for which acceleration tends to zero for
asymptotically large times. In fact, this system is found to decouple
into two independent subsystems, one parallel to and the other
orthogonal to the magnetic field. In the former subsystem, acceleration
remains time independent and corresponds to unrenormalized mass, while
damping and renormalized mass (or renormalized external force) emerges
in the latter subsystem.

It is therefore an important open question on what grounds the decrease
of the external force can be interpreted as mass renormalization, and
whether in other systems this renormalization is not only direction
dependent but, for example, also (spacetime) position and velocity
dependent (as suggested by some preliminary considerations not detailed
here).

The three methods we proposed and investigated led to the same result
for the two systems considered. Some differences among the three
approaches were found, though. First, the problematic aspect is that
there is an encoded amount of ambiguity in the partial differential
equation to solve. This ambiguity was easy to rule out for the two
systems we considered but may be a harder task for other systems.
Second, in the two iteration approaches, convergence remained a tough
open mathematical problem; moreover, not all coefficients of the model
ensured the existence of a solution in one of the iteration methods.

Further study is needed, accordingly, about each method separately and
also about some possible connections among them.

\appendix

\section{The proof utilized in section~\ref{qgy}}

The claim that \eqref{A3} yields $\LLA_3 = \LLnull$ can be proved as
follows.

We start with deriving a simple algebraic fact. Let $\LLT$ be a
symmetric trilinear map. Then, with the simplifying notation used
earlier,
 \begin{equation}  \label{tri}
\LLT \bigl( (\LLLv + \LLv)^3 \bigr) = \LLT \big( \LLLv^3 \big) + 3\LLT
\big( \LLLv^2, \LLv \big) + 3 \LLT \big(\LLLv, \LLv^2 \big) + \LLT
\big(\LLv^3 \big)
 \end{equation}
for all $\LLv$ and $\LLw$.  Thus, if $\LLT \bigl( (\Lcdot)^3 \bigr) =
\LLnull$ then $3 \LLT \big( \LLLv^2, \LLv \big) + 3 \LLT \big( \LLLv,
\LLv^2 \big) = \LLnull$ for all $\LLv$ and $\LLLv$. Here, for a fixed
$\LLv$, the first term is bilinear, the second term is linear in
$\LLLv$; their sum can be zero only if both are zero. As a consequence,
if $ \LLT \big( \LLv^3 \big) = \LLnull$ for all $\LLv$ then $\LLT
(\LLLv^2, \LLv) = \LLnull$ for all $\LLLv$ and $\LLv$. Further, for a
fixed $\LLv$, we have
 \begin{equation}
\LLT \bigl( (\LLLv_1 + \LLLv_2)^2, \LLv \bigr) = \LLT \big(\LLLv_1^2,
\LLv \big) + 2 \LLT (\LLLv_1, \LLLv_2, \LLv) + \LLT \big( \LLLv_2^2,
\LLv \big) ,
 \end{equation}
which shows that, if $\LLT \big( \LLLv^2, \LLv \big) = \LLnull$ for all
$\LLLv$ and $\LLv$, then
 \begin{equation}  \label{wwv}
\LLT (\LLLv_1, \LLLv_2, \LLv) = \LLnull
 \quad \hbox{for all} \quad
\LLLv_1, \LLLv_2, \LLv .
 \end{equation}
Hence, we have the result: if $ \LLT (\LLv^3) = \LLnull$ for all
$\LLv$ then \eqref{wwv} holds.

The previous consideration and \eqref{A3} yield that
 \begin{equation}
\bigl[ \LLI - \LKeta (\LLHB + \LLA_1) \bigr] \LLA_3 (\LLLv_1, \LLLv_2,
\LLv) = 3 \LKeta \LLA_3 \big( \LLLv_1, \LLLv_2, (\LLHB + \LLA_1) \LLv
\big)
 \end{equation}
for all $\LLLv_1$, $\LLLv_2$ and $\LLv$.
For fixed $\LLLv_1$ and $\LLLv_2$, $\LLW := \LLA_3 (\LLLv_1, \LLLv_2,
\Lcdot )$ is a linear map satisfying
 \begin{equation}  \label{llw1}
\big[ \LLI - \LKeta (\LLHB + \LLA_1) \big] \LLW = 3 \LKeta \LLW (\LLHB +
\LLA_1) ,
 \end{equation}
which implies
 \begin{equation}  \label{llw2}
\LLW \bigl[ \LLI - 3 \LKeta (\LLHB + \LLA_1) \bigr] = (\LLHB + \LLA_1)
\LKeta \LLW ,
 \end{equation}
too. Recall that $\LLHB + \LLA_1 = (1 - \Lbet) \LLHB - \Lalp\LLP$. Then
it is a simple fact that the linear maps multiplying $\LLW$ on the left
hand side in \eqref{llw1} and \eqref{llw2}, respectively, are
nondegenerate. Therefore, applying $\LLI -\LLP$ from the right to
\eqref{llw1} and from the left to \eqref{llw2}, we find
% \begin{equation}
$
\LLW \LLP = \LLW = \LLP \LLW .
$
% \end{equation}

As a consequence, $\LLW$ commutes with $\LLHB$, too, so it must be of
the form
% \begin{equation}  \label{zza}
 $
\LLW = \LLLlam \LLHB + \Lpi \LLP ,
 $
% \end{equation}
and either \eqref{llw1} or \eqref{llw2} implies
 \begin{equation}  \label{zzb}
(1 + 4\LKeta\Lalp) \LLW = 4 \LKeta (1 - \Lbet) \LLW \LLHB .
 \end{equation}
This equation and % \eqref{zza}
 $
\LLW = \LLLlam \LLHB + \Lpi \LLP
 $
result in
 \begin{align}  \label{qgu}
(1 + 4\LKeta\Lalp) \LLLlam
 &
=
% &
4 \LKeta (1 - \Lbet) \Lpi,
% \\   \label{qgv}
&
(1 + 4\LKeta\Lalp) \Lpi
 &
=
% &
- 4 \LKeta (1 - \Lbet) \LHb^2 \LLLlam ,
 \end{align}
implying $\LHb^2\LLLlam^2 = - \Lpi^2$, which is possible only if
$\LLLlam=0$ and $\Lpi=0$, i.e., $\LLW = \LLnull$.

Since we had $ \LLW = \LLA_3 (\LLLv_1, \LLLv_2, \Lcdot) $ for arbitrary
$\LLLv_1$ and $\LLLv_2$, we arrive at
% \begin{equation}
 $
\LLA_3 = \LLnull .
 $
% \end{equation}

\section*{Acknowledgments}

Support from the Hungarian Scientific Research Fund (OTKA, Grant No.
K116375) is appreciated.

\end{document}